\documentclass[aps,prl,twocolumn,superscriptaddress,amsmath,amssymb,showpacs]{revtex4}

\usepackage{graphicx}
\usepackage{dcolumn}
\usepackage{bm}
\usepackage{color}

\begin{document}

\title{Odd-Parity Pairing and Topological Superconductivity in a
Strongly Spin-Orbit Coupled Semiconductor}

\author{Satoshi Sasaki}
\author{Zhi Ren}
\author{A. A. Taskin}
\author{Kouji Segawa}
\affiliation{Institute of Scientific and Industrial Research,
Osaka University, Ibaraki, Osaka 567-0047, Japan}

\author{Liang Fu}
\email{liangfu@mit.edu}
\affiliation{Department of Physics, Massachusetts Institute of 
Technology, Cambridge, MA 02139, USA}

\author{Yoichi Ando}
\email{y_ando@sanken.osaka-u.ac.jp}
\affiliation{Institute of Scientific and Industrial Research,
Osaka University, Ibaraki, Osaka 567-0047, Japan}

\date{\today}

\begin{abstract}

The existence of topological superconductors preserving time-reversal
symmetry was recently predicted, and they are expected to provide a
solid-state realization of itinerant massless Majorana fermions and a
route to topological quantum computation. Their first concrete example,
Cu$_x$Bi$_2$Se$_3$, was discovered last year, but the search for new
materials has so far been hindered by the lack of guiding principle.
Here, we report point-contact spectroscopy experiments showing that the
low-carrier-density superconductor Sn$_{1-x}$In$_x$Te is accompanied
with surface Andreev bound states which, with the help of theoretical
analysis, give evidence for odd-parity pairing and topological
superconductivity. The present and previous finding of topological
superconductivity in Sn$_{1-x}$In$_x$Te and Cu$_x$Bi$_2$Se$_3$
demonstrates that odd-parity pairing favored by strong
spin-orbit-coupling is a common underlying mechanism for materializing
topological superconductivity.

\end{abstract}




\maketitle

Topological superconductors (TSCs) have become a research frontier in
the study of topologically ordered electronic states of matter
\cite{Schnyder, Volovik, Qi, Fu}. As a superconducting (SC) cousin of
topological insulators \cite{hasankane, zhang}, a TSC supports gapless
surface quasiparticle states consisting of massless Majorana fermions 
as its distinctive characteristic. Majorana
fermions are peculiar in that particles are their own antiparticles
\cite{Wilczek}, and they are currently attracting significant interest
because of their potential for fault-tolerant topological quantum
computing \cite{Alicea}. The $p$-wave superconductor Sr$_2$RuO$_4$ has
been widely discussed \cite{Maeno} to be an example of a chiral TSC
associated with spontaneous time-reversal symmetry breaking \cite{read}.
More recently, time-reversal-invariant TSCs were theorized and attracted
much attention \cite{zhang}. Lately Cu$_x$Bi$_2$Se$_3$ superconductor
\cite{Hor} has been theoretically proposed \cite{Fu} and experimentally
identified \cite{Sasaki} as the first material realization of such a
TSC. However, Cu$_x$Bi$_2$Se$_3$ crystals are intrinsically
inhomogeneous \cite{Kriener} and it has been difficult to elucidate the
nature of the surface Majorana fermions. Naturally, discoveries of new
TSC materials are strongly called for. In this context,
Cu$_x$Bi$_2$Se$_3$ is peculiar in that it is a superconductor obtained
by doping a topological insulator, and such materials are few and far
between; consequently, the prospect of finding new TSC materials in
doped topological insulators is not very bright. 

Nevertheless, the discovery of topological superconductivity in
Cu$_x$Bi$_2$Se$_3$ suggested that other TSCs might also be found in
low-carrier-density semiconductors whose Fermi surface is centered
around time-reversal-invariant momenta \cite{Hsieh-Fu, Qi-Fu, Michaeli}. This
motivated us to look for signatures of topological superconductivity in
In-doped SnTe (denoted Sn$_{1-x}$In$_x$Te) \cite{Bushmarina,Fisher}
whose Fermi surface depicted in Fig. 1(a) satisfies the above criteria.
In this Letter, by performing point-contact spectroscopy on
Sn$_{1-x}$In$_x$Te single crystals, we elucidate the existence of a
surface Andreev bound state (ABS) which is a hallmark of an
unconventional superconductivity \cite{Kashiwaya}. Knowing that the
symmetry and low-energy physics of this material \cite{TCI2} allows only
three types of superconducting gap functions and that all possible
unconventional states are topological, it is possible to
conclude that Sn$_{1-x}$In$_x$Te is a TSC. This discovery not only
enriches the family of TSC materials for their detailed investigations,
but also points to a common mechanism for topological superconductivity,
providing a guiding principle for the search of TSCs.

It is known that In in Sn$_{1-x}$In$_x$Te acts as an acceptor and
suppresses the ferroelectric structural phase transition (SPT) in SnTe.
Above $x \simeq$ 0.04 the SPT is completely suppressed and the system
becomes a robust superconductor whose $T_c$ gradually increases with $x$
up to $\sim$2 K at $x \simeq$ 0.10 \cite{Fisher}.
Specific-heat measurements have confirmed bulk superconductivity with 
possibly a strong pairing interaction for $x$ = 0.044 (where $T_c$ = 1.0
K) \cite{Fisher}, but no experiment to detect the surface ABS has been
carried out so far. In this Letter, we focus on samples with $x$ = 0.045
to avoid complications associated with the SPT.

\begin{figure}
\includegraphics*[width=8.7cm]{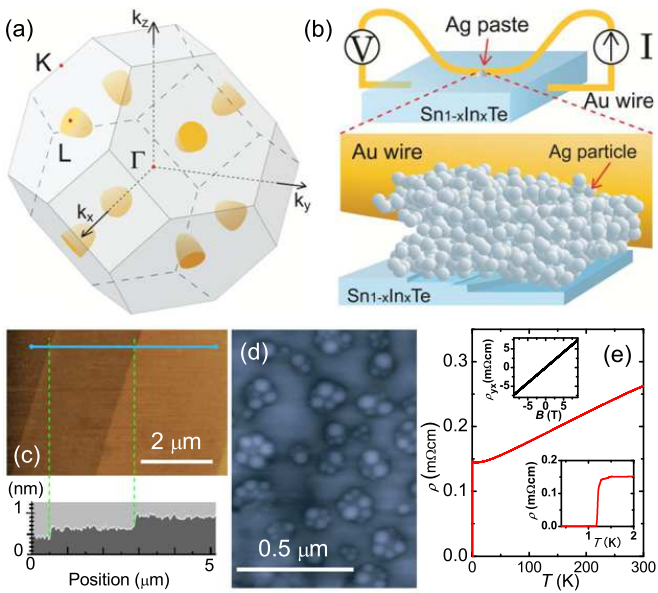}
\caption{(Color online) 
SnTe and the soft point-contact spectroscopy.
(a) The Fermi surfaces of $p$-type SnTe are centered around the four
equivalent $L$ points, which belong to the time-reversal-invariant momenta, 
in the bulk Brillouin zone of the cubic NaCl structure with FCC Bravais lattice.
(b) Schematic picture of the soft point-contact spectroscopy
experiment.
(c) Atomic-force-microscope (AFM) picture of an as-grown faceted
surface of Sn$_{1-x}$In$_x$Te single crystal and its height profile.
(d) AFM picture of the Ag particles on the measured surface after 
the gold wire is removed.
(e) Temperature dependence of the resistivity of the measured
sample ($x$ = 0.045), showing no sign of structural phase transition. 
Lower inset shows a magnified view near the SC transition at 1.2 K.
Upper inset shows the magnetic-field dependence of the Hall 
resistivity $\rho_{yx}$ showing completely $B$-linear behavior, which 
indicates that the second valence band maxima, even if slightly populated, 
plays little role in our sample; the slope gives the carrier density of 
8 $\times$ 10$^{20}$ cm$^{-3}$.
} 
\label{Fig1}
\end{figure}

SnTe crystallizes in the rock-salt structure (space group
$Fm\overline{3}m$) and hence possesses the $O_h$ point-group
symmetry. Our In-doped single crystals were grown by a vapor
transport method. High-purity elements of Sn (99.99\%), Te (99.999\%),
and In (99.99\%) were used as starting materials. The In concentration
was measured with inductively-coupled plasma atomic emission
spectroscopy (ICP-AES) and was confirmed to be consistent with the
observed $T_c$ \cite{Fisher}. The crystallographic orientation of the
surface plane was confirm by the x-ray Laue analysis to be (001). The
resistivity and the Hall resistivity were measured in the Hall-bar
geometry with a six-probe method on the same crystal [Fig. 1(e)]. The
Quantum Design PPMS was used as a platform to cool the samples down to
0.37 K and apply magnetic fields up to 9 T. The upper critical field
$H_{c2}$ defined by a sharp resistivity onset was 0.3 T at 0.37 K [Fig.
2(e)].

We performed conductance spectroscopy on the faceted (001) as-grown
surface [Fig. 1(c)] of Sn$_{1-x}$In$_x$Te single crystals with $x$ =
0.045 [$T_c$ = 1.2 K, see Fig. 1(e)] using a soft point-contact
technique \cite{Gonnelli} [Fig. 1(b)] which was successfully applied to
Cu$_x$Bi$_2$Se$_3$ \cite{Sasaki} to reveal its TSC nature. The soft
point contacts were prepared by putting a tiny drop of silver paste
below a 30-$\mu$m-diameter gold wire [Fig. 1(b)]; an atomic force
microscope image of the silver nanoparticles on a measured surface is
shown in Fig. 1(d). The $dI/dV$ spectra were measured with a lock-in
technique by sweeping a dc current that is superimposed with a
small-amplitude [1.35 $\mu$A (rms), corresponding to 0.5 A/cm$^2$] ac
current, and a quasi-four-probe configuration was employed to read the
voltage between a normal metal (silver paste) and the sample (see Ref.
\cite{Sasaki} for details). We show in the Supplemental Material
\cite{SM} that this technique yields ordinary Andreev reflection spectra
\cite{BTK} when applied to the conventional $s$-wave superconductor Sn. 

\begin{figure}
\includegraphics*[width=8.7cm]{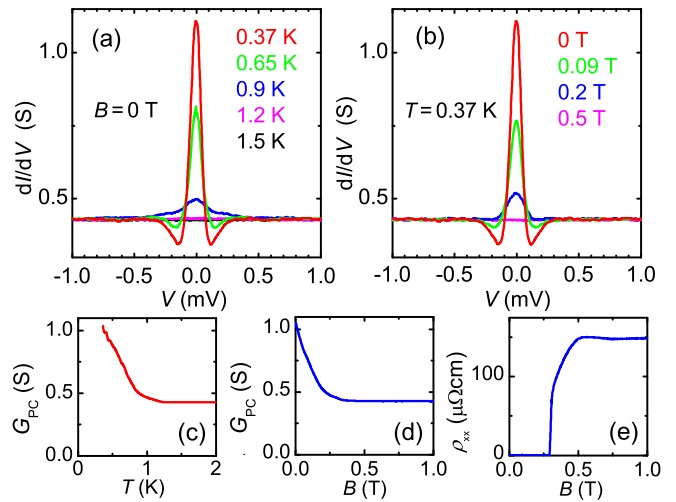}
\caption{(Color online) 
Conductance spectra of Sn$_{1-x}$In$_x$Te.
(a) Bias-voltage ($V$) dependence of the differential conductance,
$\frac{dI}{dV}(V)$, at various temperatures in 0 T.
(b) $\frac{dI}{dV}(V)$ in various magnetic field
at 0.37 K.
(c) Temperature dependence of the zero-bias conductance in 0 T, showing its
enhancement of more than a factor of two in the SC state at low temperature.
(d) Magnetic-field ($B$) dependence of the zero-bias conductance at 0.37 K.
(e) $B$ dependence of the resistivity $\rho_{xx}$ at 0.37 K showing
that 0.5 T is enough to completely suppress superconductivity.
} 
\label{Fig2}
\end{figure}

When applied to Sn$_{1-x}$In$_x$Te, this technique allowed us to observe
a signature of ABS \cite{Tanaka} [Figs. 2(a)-2(b)], rather than the
ordinary Andreev reflection; namely, the bias-voltage dependence of the
differential conductance $dI/dV$ presents a pronounced peak at zero
voltage (i.e. Fermi level) accompanied by dips on its sides at the
energy scale of the SC gap ($\pm$0.1 meV). In the case of the ordinary
Andreev reflection \cite{BTK}, as one can see in Fig. S1 of the
Supplemental Material \cite{SM}, two peaks, rather than dips, should be
observed at the SC gap energy at low enough temperatures. Moreover, in
our data for Sn$_{1-x}$In$_x$Te, the point-contact conductance at zero
energy, $G_{\rm PC}$, in the SC state becomes more than twice the
normal-state value [Figs. 2(c)-2(d)], which is impossible for Andreev
reflections \cite{BTK} and points to the existence of ABS on the surface
\cite{Kashiwaya,Tanaka}. 

The large magnitude of the observed zero-bias conductance peak (ZBCP) is
already a strong indication that it is due to an ABS, but it is prudent
to examine possible relevance of other origins of the ZBCP, such as
heating effect \cite{Sheet}, reflectionless tunneling \cite{Beenakker},
and magnetic Kondo scattering \cite{Rowell}. In this respect, the
magnetic-field dependence of the spectra [Fig. 2(b)] gives evidence
against those other possibilities (see Supplemental Material \cite{SM}
for details) and one can conclude with confidence that the observed ZBCP
is caused by an inherent surface ABS. This gives strong evidence for an
unconventional SC state in Sn$_{1-x}$In$_x$Te. 

To identify the nature of the SC state in Sn$_{1-x}$In$_x$Te, we first note 
that the Fermi surface in the normal state consists
of four ellipsoids centered at four $L$ points of the face-centered-cubic
(FCC) Brillouin zone. The conduction and
valence bands in the vicinity of each $L$ point are described by the
$k\cdot p$ Hamiltonian \cite{TCI2}:
\begin{eqnarray}
H({\bf k})= m \sigma_z + v \sigma_x (k_1 s_2 - k_2 s_1) + v_3 k_3 \sigma_y .
\label{H}
\end{eqnarray} 
Here $k_3$ is the momentum along the three-fold axis $\Gamma L$; $k_2$
is along the two-fold axis $LK$. $s_i$ and $\sigma_i$ are Pauli matrices
associated with spin and orbital degrees of freedom, respectively.
Specifically, the two orbitals labeled by $\sigma_z = \pm 1$ are mainly
derived from the $p$-orbitals of Sn and Te atoms, respectively. We
emphasize that at $L$ points, these two types of $p$-orbitals have
opposite parity and do not mix. The four-band Hamiltonian (\ref{H}) of
Sn$_{1-x}$In$_x$Te at the $L$ points of the FCC lattice is essentially
equivalent to that of Cu$_x$Bi$_2$Se$_3$ at the $\Gamma$ point of the
rhombohedral lattice \cite{Hsieh-Fu}, both of which are dictated by the
underlying $D_{3d}$ point group symmetry. 

We now discuss the possible pairing symmetries. Since the four $L$
points are invariant under the inversion of crystal momentum ${\bf k}
\rightarrow -\bf k$, superconducting order parameters with zero total
momentum correspond to pairing {\it within} each Fermi pocket, and
therefore consist of four components on the four Fermi pockets: ${\vec
\Delta}= (\Delta_1, \Delta_2, \Delta_3, \Delta_4)$. Each $\Delta_j$ can
be classified by the representations of $D_{3d}$, a subgroup of the
$O_h$ point group for In-doped SnTe that leaves $L_j$ invariant. For the
Hamiltonian (\ref{H}) at a given $L_j$, there are four types
of momentum-independent gap functions $\Delta_j$ with different internal
spin and orbital structures, corresponding to the $A_{1g}, A_{1u}$,
$A_{2u}$ and $E_u$ representations of $D_{3d}$ \cite{Fu}. Furthermore,
depending on the relative phases between $\Delta_{1}, ...., \Delta_4$,
$\vec \Delta$ belong to different representations of the $O_h$ point
group. It is beyond the scope of this paper to exhaust all
possibilities. Instead, we consider those superconducting states that do
not spontaneously break any lattice symmetry, in accordance with all
experimental facts known so far. There are three such states
corresponding to the following one-dimensional representations of $O_h$
point group: $A_{1g},A_{1u}$ and $A_{2u}$. (The $E_u$ state breaks the
three-fold rotation symmetry around (111) axis.)

Among these three states, $A_{1g}$ is even-parity and fully-gapped,
which corresponds to an $s$-wave superconductor and do not have a surface
ABS. Both $A_{1u}$ and $A_{2u}$ states are unconventional
superconductors with odd-parity pairing. The $A_{1u}$ state is
fully-gapped, and realizes an odd-parity TSC. The topological invariant
is given by $N=\sum_j N_j$, where $|N_j|=1$ is the invariant associated
with each Fermi surface and its sign is given by ${\rm sign}
(\Delta_j)$ \cite{Qi-Fu}. Importantly, the four components $\Delta_1, ...
\Delta_4$ are related by rotation symmetry and have the same sign in the
$A_{1u}$ state. As a result, the $A_{1u}$ state of In-doped SnTe is a
TSC with $|N|=4$, which supports topologically protected surface ABS. 

The odd-parity $A_{2u}$ state has point nodes at the intersection of
each Fermi pocket with the $\Gamma L$ line. These nodes are protected by
the mirror symmetry of the FCC crystal structure. While it is impossible
to define a 3D topological invariant for a gapless phase, one can still
define ``weak" topological invariants associated 2D
time-reversal-invariant planes in the Brillouin zone \cite{moore, fkm}
that are fully-gapped. For the $A_{2u}$ state in In-doped SnTe, any
plane that passes a {\it single} $L$ point and avoids the nodes
satisfies the criterion for 2D odd-parity TSC \cite{Fu, Sato} and has a
nonzero weak topological invariant. As a result, the $A_{2u}$ state has
topologically protected ABS, similar to those in the $A_{2u}$ state of 
Cu$_{x}$Bi$_2$Se$_3$ theoretically demonstrated earlier \cite{Sasaki}.

From the above analysis, we conclude that the two odd-parity states
$A_{1u}$ and $A_{2u}$ are topologically nontrivial and support ABS that
can naturally give rise to the observed ZBCP. We further propose an
electron-phonon mechanism for odd-parity pairing in Sn$_{1-x}$In$_x$Te.
First, we note that SnTe has a soft transverse optical (TO) phonon at
$\bf q=0$, which couples strongly to {\it interband} electronic
excitations \cite{Sugai}. This phonon mode corresponds to the
displacement of Sn and Te sublattices relative to each other. It becomes
unstable and leads to the SPT at low temperature. The SPT temperature is
suppressed by In doping \cite{Fisher}. As one can see in Fig. 1(e), the
temperature dependence of the resistivity shows no kink down to $T_c$,
which indicates that the SPT is completely suppressed in our sample;
according to the phase diagram \cite{Fisher}, this is reasonable for $x$
= 0.045. This suggests that the TO phonon remains stable. Moreover,
proximity to the SPT suggests that the tendency toward ferroelectricity
is strong, which naturally points to an attractive interaction between
Sn and Te $p$ orbitals. 

Assuming that such an {\it inter-orbital} attraction from
electron-phonon coupling is the origin for superconductivity, we can now
theoretically deduce the pairing symmetry of Sn$_{1-x}$In$_x$Te by
following a similar analysis as was done for Cu$_{x}$Bi$_2$Se$_3$
\cite{Fu}. Essentially, the spin-orbit-coupled band structure (\ref{H})
cooperates with the above attractive interaction to favor the pairing
between Sn and Te orbitals. Because the two orbitals have opposite
parity as mentioned earlier, one may conclude that the pairing symmetry
in Sn$_{1-x}$In$_x$Te is most likely odd-parity. A detailed theory of
the pairing mechanism and a full determination of the type of odd-parity
pairing is beyond the scope of this paper and will be presented
elsewhere \cite{forthcoming}. In any case, since the even-parity state
does not produce a surface ABS but both odd-parity states in
Sn$_{1-x}$In$_x$Te do, our experimental observation indicates that the
odd-parity pairing is realized in this material, which agrees with the
above theoretical consideration for the pairing mechanism. Given that
the two possible odd-parity states are both topological as already
discussed, the present observation leads to the conclusion that
Sn$_{1-x}$In$_x$Te is a TSC. Note that the similarities in both band
structure and pairing symmetry between Cu$_x$Bi$_2$Se$_3$ and
Sn$_{1-x}$In$_x$Te naturally suggest that the calculations of the
surface ABS performed for Cu$_x$Bi$_2$Se$_3$ \cite{Sasaki, Hsieh-Fu,
Yamakage, Lee} should also hold qualitatively for Sn$_{1-x}$In$_x$Te.
This means that the observed ZBCP is exactly what is theoretically
expected for this type of TSC. In passing, it is useful to note that in
the conductance spectra of previously known TSCs (Sr$_2$RuO$_4$
\cite{Kashiwaya2011} and Cu$_x$Bi$_2$Se$_3$ \cite{Sasaki}), the unusual
dip at the SC gap energy scale has been consistently observed, bearing
strong similarly to the present spectra.

\begin{figure}
\includegraphics*[width=8.7cm]{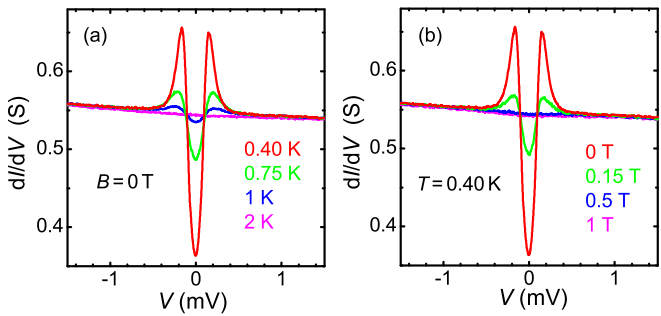}
\caption{(Color online) 
Conductance spectra of Pb$_{1-x}$Tl$_x$Te.
(a) Bias-voltage ($V$) dependence of the differential conductance,
$\frac{dI}{dV}(V)$, at various temperatures in 0 T.
(b) $\frac{dI}{dV}(V)$ in various magnetic fields
at 0.40 K. The sample ($x$ = 0.013) measured here had the zero-resistivity
$T_c$ of 1.35 K and the carrier density of 1.3 $\times$ 10$^{20}$ cm$^{-3}$.
} 
\label{Fig3}
\end{figure}

Finally, we have also studied a similar superconducting material
Pb$_{1-x}$Tl$_x$Te with the same point-contact technique
\cite{PbTe_note}. Both Sn$_{1-x}$In$_x$Te and Pb$_{1-x}$Tl$_x$Te
superconduct at 1 -- 2 K \cite{FisherPbTe}, crystallize in rock-salt
structure, and have similar band structures. Nevertheless, we found no
evidence for a TSC state in Pb$_{1-x}$Tl$_x$Te (Fig. 3); the observed
data is consistent with the Andreev reflection spectra of conventional
SC state with a low contact transparency (close to the tunneling limit).
The occurrence of the conventional SC state here is probably because the
pairing interaction in Pb$_{1-x}$Tl$_x$Te is dominated by the charge
Kondo mechanism \cite{FisherPbTe} which leads to ordinary $s$-wave
pairing. This comparison seems to indicate the importance of the TO
phonon for TSC in this class of materials.

The discovery of topological superconductivity in Sn$_{1-x}$In$_x$Te
reported here is instructive for further explorations of new TSCs: It
gives us a guiding principle to look for semiconductors with strong
spin-orbit coupling and having Fermi surfaces surrounding
time-reversal-invariant momenta, because the occurrence of TSC in both
Cu$_x$Bi$_2$Se$_3$ and Sn$_{1-x}$In$_x$Te strongly suggests a common
mechanism. In addition, this discovery has a practical importance: While
the previously discovered TSC material Cu$_x$Bi$_2$Se$_3$ suffers a
problem of intrinsic inhomogeneity \cite{Kriener} which hindered detailed
studies, high-quality single crystals of Sn$_{1-x}$In$_x$Te with 100\%
SC volume fraction are readily available. Hence, Sn$_{1-x}$In$_x$Te
makes it possible to explore the new topological state of matter, the
time-reversal-invariant TSC, on a robust platform for the first time.

\begin{acknowledgments} 

We thank T. Ueyama and R. Yoshida for technical assistance. 
This work was supported by JSPS (NEXT Program and KAKENHI 24740237), 
MEXT (Innovative Area ``Topological Quantum Phenomena" KAKENHI), 
and AFOSR (AOARD 124038).

\end{acknowledgments}

\clearpage
\onecolumngrid

\renewcommand{\thefigure}{S\arabic{figure}} 

\setcounter{figure}{0}

\renewcommand{\thesection}{S\arabic{section}.} 

\begin{flushleft} 
{\Large {\bf Supplemental Material}}
\end{flushleft} 

\vspace{2mm}

\begin{flushleft} 
{\bf S1. Soft Point-Contact Spectra on a Conventional Superconductor}
\end{flushleft} 

To confirm that our soft point-contact technique yields conventional
Andreev reflection spectra if applied to a conventional
superconductor, we have prepared a soft point contact on an oxidized 
surface of Sn with the same technique using the same silver paint as the 
experiment for Sn$_{1-x}$In$_x$Te. As shown in Fig. S1, the obtained 
spectra are what is exactly expected for conventional Andreev reflection 
between a BCS superconductor and a metal. The fitting of the extended 
Blonder-Tinkham-Klapwijk theory \cite{Gonnelli} to the data tells us 
that this junction has an intermediate transparency of $Z$ = 0.47.

\begin{figure}[b]
\begin{center}
\includegraphics[clip,width=6cm]{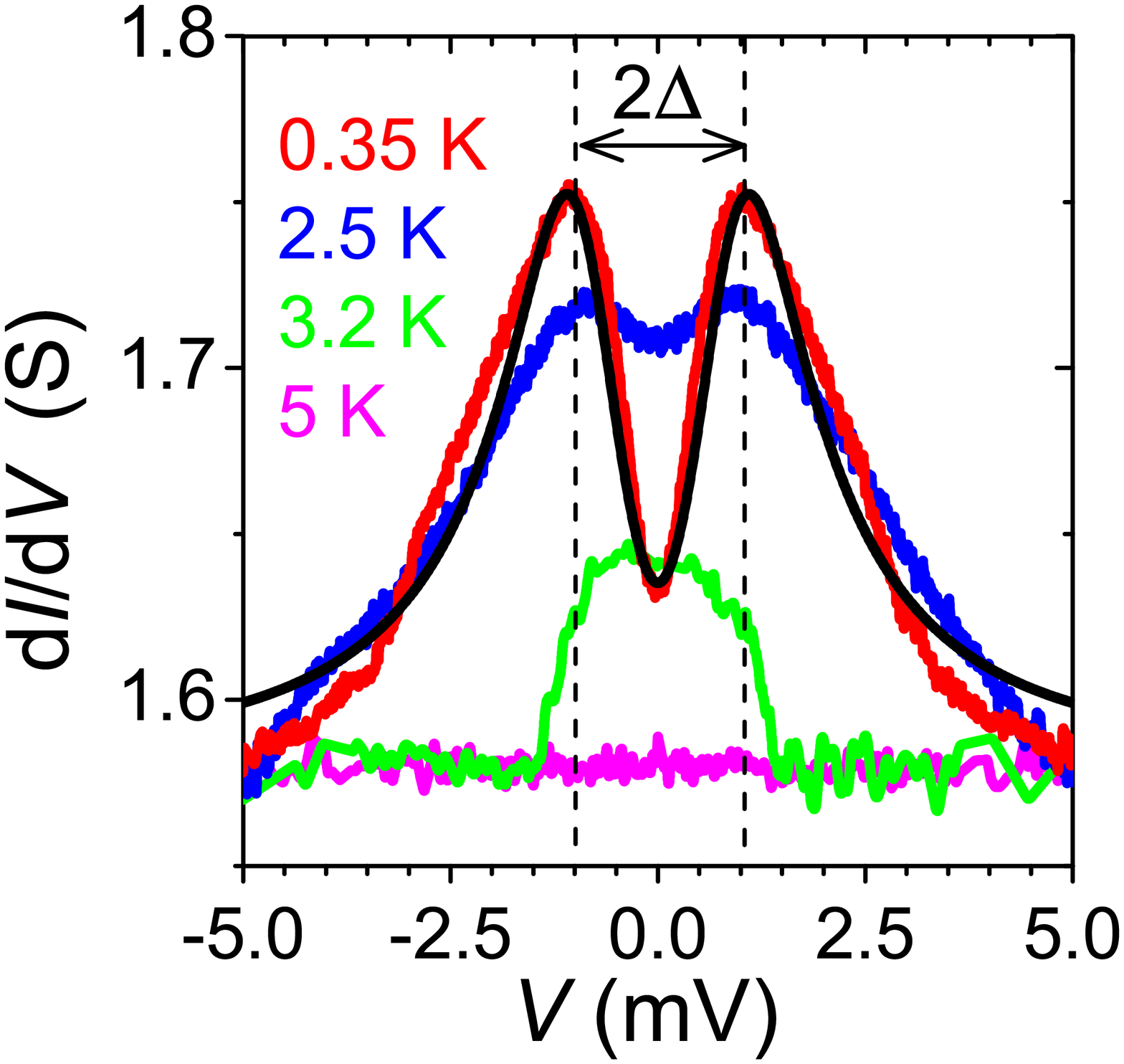}
\caption{Andreev reflection spectra measured with the soft point contact 
technique on a conventional superconductor. Differential conductivity 
$dI/dV$ vs. bias voltage obtained on a oxidized surface of Sn with our 
soft point-contact technique. The black solid line is the theoretical 
fitting to the data at 0.35 K using the extended Blonder-Tinkham-Klapwijk 
theory, which yields the superconducting gap energy $\Delta$ = 1.0 meV, 
junction transparency $Z$ = 0.47, and the broadening parameter 
$\Gamma$ = 0.75 meV.
} 
\end{center}
\end{figure}

\begin{flushleft} 
{\bf S2. Origin of the Zero-Bias Conductance Peak}
\end{flushleft} 

The zero-bias conductance peak (ZBCP) is often observed in point contact
experiments on bulk superconductors. Besides the intrinsic Andreev bound
state associated with unconventional superconductivity, there are other
possible causes of the ZBCP: conventional Andreev reflection
\cite{Gonnelli,BTK}, heating/critical-current effects \cite{Sheet},
reflectionless tunneling \cite{Beenakker}, and magnetic Kondo scattering
\cite{Rowell,Appelbaum}. In the following, we show how those other causes 
of the ZBCP can be dismissed.

The conventional Andreev reflection can give rise to a single ZBCP when
the temperature is close to $T_c$. This is because the two-peak
structure characteristic of the conventional Andreev reflection spectra
can be blurred into a single peak by thermal fluctuations near $T_c$.
For example, in our data for Sn shown in Fig. S1, one can see a single
peak in the data for 3.2 K, which is only 0.5 K below $T_c$; as the
temperature is lowered, such a single peak splits into two, and the peak
separation corresponds to twice the superconducting energy gap,
$2\Delta$. Indeed, in the case of Sn, the peak splits into two at 2.5 K
and below, and the peak separation is consistent with $2\Delta$. On the
other hand, because the conductance peak in Sn$_{1-x}$In$_x$Te does not
split into two even at 0.37 K (which is less than 1/3 of $T_c$) and also 
because pronounced dips rather than peaks are observed at $2\Delta$, one 
can safely conclude that the spectra observed in Sn$_{1-x}$In$_x$Te is not
due to the conventional Andreev reflection.

The heating/critical-current effects cause a spurious ZBCP in samples
with a large normal-state resistivity when the contact is in the thermal
regime. This is because the increase in the bias voltage causes the
local current to exceed the critical current and leads to a
voltage-dependent decrease in the differential conductivity. A
characteristic feature of the ZBCP of this origin is that the height of
the peak changes little with a weak magnetic field, while the width of
the ZBCP narrows quickly with increasing $H$. This is because the
conductivity at zero-bias is always measured below the critical current
as long as the superconductor is in the zero-resistivity state, whereas
the decrease in critical current with magnetic field is directly
reflected in the width of the spurious ZBCP. In the magnetic-field
dependence of the spectra measured on Sn$_{1-x}$In$_x$Te [Fig. 2(b) of
the main text], the ZBCP is suppressed to about a half of the height
at 0 T with a magnetic field of only 0.09 T, but the position of
the dips in the spectra does not move inward. Obviously, this speaks
against the heating/critical-current origin of the ZBCP. Also, when the
critical current is reached in a point contact, it usually causes a
sharp, spike-like dip in the $dI/dV$ data \cite{Gonnelli,Sheet},
which reflects a rapid increase in voltage with a small change in
current. The absence of such spikes in our data corroborates our
conclusion that the heating/critical-current effect is not relevant to
the observed ZBCP.

In the case of the reflectionless tunneling \cite{Beenakker}, the
conductance peak is caused by the coherent backscattering of electrons
in the normal metal. Such a coherent backscattering is suppressed with a
magnetic field because it breaks time-reversal symmetry, and the
characteristic magnetic field for the suppression is estimated with the
condition that the magnetic flux threading a coherent path of the
electrons (which is set either by the size of the normal metal or by the
electron dephasing length $L_{\phi}$) is equal to the flux quantum
$\Phi_0$ ($= h/2e$). The footprint of our point contact has the diameter
of $\sim$20 $\mu$m, which is larger than the dephasing length $L_{\phi}$
of $\sim$1 $\mu$m in Ag below 1 K \cite{Lphi}. This means that
$L_{\phi}$ governs the reflectionless tunneling \cite{Beenakker} which
would be suppressed with only $\sim$0.7 mT. Given that the ZBCP is still
visible in 0.2 T in the present case, the reflectionless tunneling is
not likely to be its origin. Also, the reflectionless tunneling is a
weak effect and can never cause a doubling of the zero-bias conductance.

Another well-known origin of the ZBCP is the magnetic Kondo scattering
\cite{Rowell,Appelbaum}. One should remember that the ZBCP of this
origin is expected even in the normal state, and, hence, the fact that
the ZBCP in Sn$_{1-x}$In$_x$Te shows up only in the superconducting
state already speaks against this possibility. Furthermore, the $g$
factor of SnTe is expected to be more than 20 for the valence band
\cite{Bernick}, which suggests that the Zeeman splitting for $S$ = 1/2
in 0.1 T would be larger than 0.1 meV. This should be observable in our
data if the magnetic scattering is relevant. Hence, the absence of peak
splitting in 0.1 T gives further evidence against the relevance of the
magnetic Kondo scattering in the ZBCP.


\begin{thebibliography}{99}
\bibitem{PbTe_note}
Pb$_{1-x}$Tl$_x$Te single crystals were grown by a vapor transport
method using high-purity elements of Pb (99.998\%), Te (99.999\%), and
Tl (99.999\%). The Tl concentration was measured with ICP-AES and was
consistent with the $T_c$ value \cite{FisherPbTe}. For the point-contact
experiments, Pb$_{1-x}$Tl$_x$Te crystals were cleaved at room
temperature in air to obtain a good (001) surface.
\end{thebibliography}
\end{document}